\documentclass{article}
\usepackage{icmc2025_paper_template}
\usepackage{times}
\usepackage{ifpdf}
\usepackage{soul}
\usepackage[english]{babel}
\usepackage[symbol]{footmisc}

\def\papertitle{Paper Template for ICMC 2025}
\def\firstauthor{Yonghyun Kim}
\def\secondauthor{Sangheon Park}
\def\thirdauthor{Marcus Parker}
\def\fourthauthor{Donghoon Seu}
\def\fifthauthor{Alexandria Smith}

\newif\ifpdf
\ifx\pdfoutput\relax
\else
   \ifcase\pdfoutput
      \pdffalse
   \else
      \pdftrue
  \fi
\fi

\ifpdf 
  \usepackage[pdftex,
    pdftitle={\papertitle},
    pdfauthor={\firstauthor, \secondauthor, \thirdauthor, \fourthauthor, \fifthauthor},
    bookmarksnumbered, 
    pdfstartview=XYZ 
   ]{hyperref}

  \usepackage[pdftex]{graphicx}
  \graphicspath{{./figures/}}
  \DeclareGraphicsExtensions{.pdf,.jpeg,.png}

  \usepackage[figure,table]{hypcap}

\else 
  \usepackage[dvips,
    bookmarksnumbered, 
    pdfstartview=XYZ 
  ]{hyperref}  

  \usepackage[dvips]{epsfig,graphicx}
  \graphicspath{{./figures/}}
  \DeclareGraphicsExtensions{.eps}

  \usepackage[figure,table]{hypcap}
\fi

\hypersetup{
    colorlinks,%
    citecolor=black,%
    filecolor=black,%
    linkcolor=black,%
    urlcolor=black
}

\title{NeoLightning: A Modern Reimagination of Gesture-Based Sound Design}

 \fiveauthors
  {0.5in}
  {\firstauthor\textsuperscript{*}} {Georgia Institute of Technology \\ %
    {\tt \href{yonghyun.kim@gatech.edu}{yonghyun.kim@gatech.edu}}}
  {\secondauthor\textsuperscript{*}} {Georgia Institute of Technology \\ %
    {\tt \href{sangheon@gatech.edu}{sangheon@gatech.edu}}}
  {\thirdauthor\textsuperscript{*}} {Georgia Institute of Technology \\ %
    {\tt \href{mparker98@gatech.edu}{mparker98@gatech.edu}}}
  {\fourthauthor} {Georgia Institute of Technology \\ %
    {\tt \href{seuseu@gatech.edu}{seuseu@gatech.edu}}}
  {\fifthauthor} {Georgia Institute of Technology \\ %
    {\tt \href{alexandria.smith@gatech.edu}{alexandria.smith@gatech.edu}}}

\makeatletter
\g@addto@macro\@author{%
  \\[4ex]
  {\noindent
  \centering
  \footnotesize
  \textsuperscript{*}These authors contributed equally to this work.\par
  }
}
\makeatother

\begin{document}
\sloppy

\capstartfalse
\maketitle
\capstarttrue
\begin{abstract}
This paper introduces NeoLightning, a modern reinterpretation of the Buchla Lightning. NeoLightning preserves the innovative spirit of Don Buchla's ``Buchla Lightning'' (introduced in the 1990s) while making its gesture-based interaction accessible to contemporary users. While the original Buchla Lightning and many other historical instruments were groundbreaking in their time, they are now largely unsupported, limiting user interaction to indirect experiences. To address this, NeoLightning leverages MediaPipe for deep learning-based gesture recognition and employs Max/MSP and Processing for real-time multimedia processing. The redesigned system offers precise, low-latency gesture recognition and immersive 3D interaction. By merging the creative spirit of the original Lightning with modern advancements, NeoLightning redefines gesture-based musical interaction, expanding possibilities for expressive performance and interactive sound design.

\end{abstract}
%


\section{Introduction}\label{sec:introduction}
There is a long-standing tradition of reviving and reimagining historical music and musical instruments that have become inaccessible or obsolete, leveraging modern technology to integrate them into contemporary contexts \cite{han2024, setiawan2024gamelan}. In recent years, the development of synthesizers and Digital Musical Instruments (DMI) has exemplified this trend, offering modern reinterpretations of classical instruments through the integration of advanced technologies \cite{glassHarmonica, Magnusson15032021, kuik2004digital}.

Reconstructing historical musical instruments that once played a pivotal role in their time provides audiences with unprecedented opportunities for direct and interactive engagement \cite{UcarMaral2023ViolaOrganista}. These reconstructions not only recreate the sounds and functionalities of the original instruments but also offer valuable insights into their cultural and historical significance. By bridging past traditions with modern technology, they deepen our understanding of the contexts that shaped these instruments and inspire new innovations in musical interaction and performance \cite{glassHarmonica}.

Beyond historical preservation,  reimagining musical instruments also fosters innovation in DMI design by incorporating modern technologies \cite{kuik2004digital}. Inspired by this approach, we set out to reimagine the Buchla Lightning, an influential DMI that pioneered gestural interaction in digital music \cite{rich1991buchla}. 

By revisiting Don Buchla's philosophy of exploring new interaction modalities, we introduce \textit{NeoLightning}, a modern reinterpretation that merges technological advancements with conventional programming tools to address the inaccessibility of the original Buchla Lightning. The project integrates MediaPipe \cite{lugaresi2019mediapipe}, a deep learning-based human landmark detector, with Max/MSP, a widely used visual programming environment for real-time audio processing. We propose a versatile, camera-based user interface designed to enhance user experience, particularly in live performance settings. This reimagined design preserves the creative intent and innovative philosophy of the original Buchla Lightning while incorporating contemporary enhancements that ensure its relevance in modern contexts. 

By bridging historical ingenuity with cutting-edge technology, \textit{NeoLightning} redefines gesture-based musical interaction, expanding the possibilities for interactive sound design and expressive music control.

\section{Background and Related Works}\label{sec:page_size}

\subsection{Buchla Lightning}

The Buchla Lightning utilized two handheld wands equipped with infrared sensors to track the performer's two-dimensional movements, enabling expressive and fluid sound control. While earlier gesture-controlled instruments, such as the Theremin and Ondes Martenot, established a connection between motion and sound, the Buchla Lightning extended this paradigm by integrating MIDI technology with advanced spatial tracking. 

\begin{figure} [h]
  \centering
  \includegraphics[width=\linewidth]{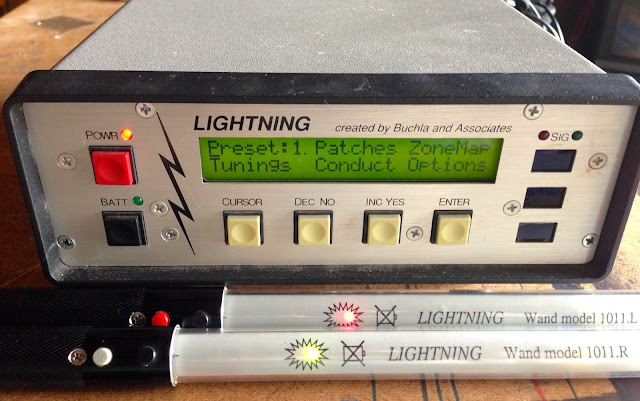}
  \caption{Buchla Lightning and Gesture-Based Instruments (Source \cite{matrixsynth2014buchla})}
\end{figure}

The Buchla Lightning series had a profound impact on gesture-based control systems and Human-Computer Interaction (HCI) in music technology \cite{casciato2008}. Its innovative design has inspired a wide range of applications, including live electronic music performances \cite{winkler1995making}, multimedia art \cite{collicutt2009real}, and even classical music performances \cite{lygesture, iSymphony}. 

Since the introduction of Buchla Lightning, numerous advancements in gesture-based musical instruments and interfaces have explored innovative approaches to bridge performer movement and sound generation. Dextoria utilized guitar-playing gestures to apply effects dynamically, eliminating the need for traditional foot pedals \cite{bernreiter2024dextoria}. MogMi demonstrated the potential of mobile devices as musical instruments by leveraging their built-in sensors to interpret gestures \cite{dekel2008mogmi}. Intonaspacio integrates spatial sound with gesture-based interaction between the performer and the instrument \cite{rodrigues2015intonaspacio}. These systems have expanded the possibilities for expressive interaction in music performance. 

\subsection{Reimagining History with Modernity}
Historical musical instruments have been successfully revived and modernized through digital music technologies, demonstrating how traditional craftsmanship and modern computational tools can coexist. A prime example is the Viola Organista, originally conceived by Leonardo da Vinci, which was meticulously reconstructed and showcased by Sławomir Zubrzycki in 2012. This project demonstrated how historical designs can be realized using modern craftsmanship and technology \cite{UcarMaral2023ViolaOrganista}.

Similarly, Välimäki, Laurson, and Erkut transformed the clavichord, a pivotal instrument of the Baroque era, into a modern synthesizer, bridging the gap between traditional acoustic instruments and contemporary digital music technologies \cite{clavichord}. Extending this approach, Paisa, Erkut, and Serafin digitized Franklin's Armonica, a rare glass instrument, by integrating it into Max/MSP and PureData, making it accessible within interactive digital music environments \cite{glassHarmonica}.

Beyond instrument reconstruction, digital technologies have been used to enhance interactive music experiences. Lee, Wolf, Jansen, and Borchers developed an innovative interactive music exhibit featuring medieval instruments such as the hurdy-gurdy, harp, and frame drum. This exhibit fostered collaborative improvisation in an e-learning space, creating unique and engaging musical experiences for audiences \cite{RexBand}.

Together, these projects highlight how the fusion of historical artifacts and digital technologies can open new frontiers in education, performance, and audience interaction, reinforcing the relevance of historical instruments in contemporary music research.

\section{Implementation of NeoLightning}\label{sec:typeset_text}
\begin{figure} [h]
  \centering
  \includegraphics[width=\linewidth]{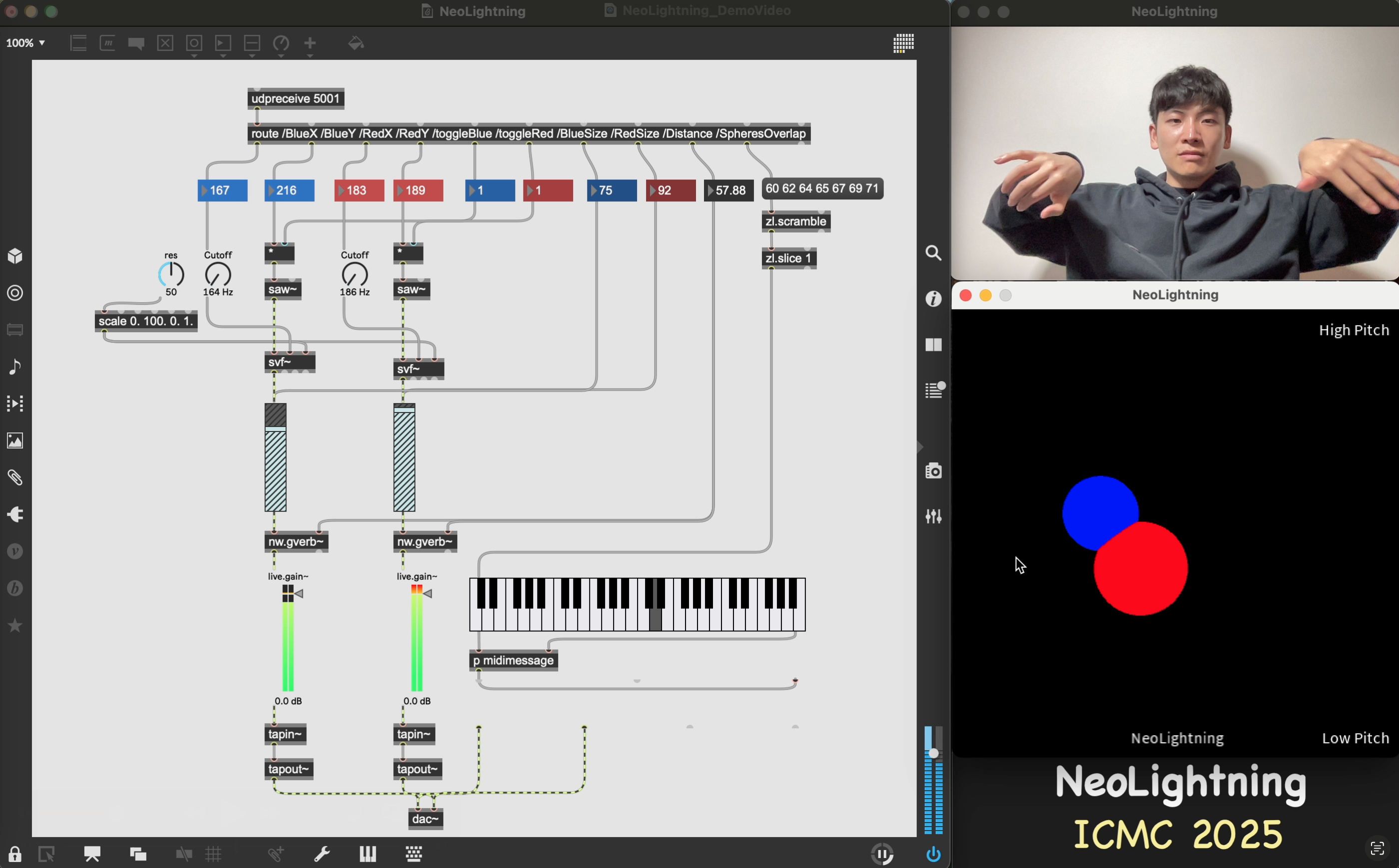}
  \caption{System workflow integrating gesture, sound, and visualization}
\end{figure}

The implementation of the system is available for reproducibility and further exploration.\footnote{Code and demo available at: \url{https://github.com/yonghyunk1m/ReimaginingTheBuchlaLightning}}

\subsection{User Interface} 
Our system simulates and enhances gesture-based control for interactive sound generation, addressing the limitations of the Buchla Lightning by leveraging MediaPipe for 3D interaction. The detected gestures are visualized in Processing \cite{processing}, which provides an intuitive platform for mapping gestures into expressive controls, allowing users to comprehend the instrument's functionality while preserving its original intent. 

The system primarily utilizes MediaPipe Hands \cite{zhang2020mediapipe}, a state-of-the-art human hand landmark detection framework, to capture and interpret real-time hand movements via a standard webcam or similar camera device. By processing human hand gestures, MediaPipe enables users to engage with the system in an intuitive and expressive manner, eliminating the need for specialized hardware required by the original Buchla Lightning. This camera-based interface serves as the primary input method, offering precise control and immediate visual feedback, allowing seamless interaction without external controllers.

Recognizing the importance of ensuring accessibility across different platforms and hardware configurations \cite{mti3030057}, the system also includes a keyboard input mode for users without access to a camera or those operating in environments where camera-based tracking is impractical. This alternative input method allows users to manually control wand positions and states, ensuring a broader range of usability. 

\subsection{Gesture Data Processing}
\begin{table}[h]
    \centering
    \renewcommand{\arraystretch}{1.2}
    \begin{tabular}{|c|c|}
        \hline
        \textbf{Keyboard / Hands} & \textbf{Action}\\ 
        \hline
        W / LH Up  & Move blue sphere up\\ 
        A / LH Left & Move blue sphere left\\ 
        S / LH Down & Move blue sphere down\\ 
        D / LH Right & Move blue sphere right\\ 
        \hline
        $\uparrow$ / RH Up & Move red sphere up\\ 
        $\leftarrow$ / RH Left & Move red sphere left\\ 
        $\downarrow$ / RH Down & Move red sphere down \\ 
        $\rightarrow$ / RH Right & Move red sphere right\\ 
        \hline
        Q / Open LH & Bring blue sphere closer\\ 
        Z / Close LH & Push blue sphere farther\\ 
        E / Open RH & Bring red sphere closer\\ 
        C / Close RH & Push red sphere farther\\ 
        \hline
        O / Swipe LH & Toggle blue sphere\\ 
        P / Swipe RH & Toggle red sphere\\ 
        \hline
    \end{tabular}
    \caption{Mapping of keyboard or hand gestures to actions in Processing}
    \label{tab:gestures_visualization_mapping}
\end{table}

The Python server employs OpenCV \cite{opencv_library} to capture image frames from the camera in real-time. These frames are then processed using MediaPipe to extract hand landmarks. To improve gesture recognition accuracy, the system incorporates hand landmarks detected in the previous frame alongside those from the current frame. The extracted gesture data is subsequently transmitted to Processing via the \texttt{udpreceive} object. In contrast, keyboard inputs are handled directly within Processing, enabling users to manually control wand positions and states without requiring external processing.

Within Processing, both gesture and keyboard input data are visualized in a simulated 3D space, generating a dynamic real-time representation (Table~\ref{tab:gestures_visualization_mapping}). This approach offers two key advantages over the Buchla Lightning. First, users gain an additional dimension for sound manipulation, as the Buchla Lightning operates within a 2D space. Second, the system provides a continuous visual representation of the interaction space, enhancing users' ability to perceive and refine their gesture-to-sound mapping. By integrating real-time feedback, the visualization allows for more precise and intuitive control over sonic parameters.

Each wand is represented as a color-coded sphere—blue for the left wand and red for the right—to ensure clear differentiation. This immediate visual feedback enhances user interaction and facilitates a deeper understanding of the system’s functionality, ultimately improving the expressivity and accuracy of gesture-based sound control.

\subsection{Sound Implementation}

\textit{NeoLightning} utilizes Max/MSP to process gesture data, translating spatial movement into musical control parameters. The audio is generated using sawtooth wave oscillators, with frequency and amplitude directly modulated by the user's hand gestures. Gesture data, which include positional coordinates and interaction states, are first routed through the \texttt{route} object for categorization. They are then scaled to map raw input values onto musically relevant control ranges, ensuring expressive and dynamic sound manipulation.

In general, the sphere's position directly interacts with the sound system, serving as a primary control mechanism for timbre, pitch, and loudness. The Y-axis position controls pitch, with lower positions corresponding to lower frequencies. The Z-axis position regulates loudness; as the sphere moves closer and increases in size, the sound intensity proportionally increases. A detailed breakdown of the sphere’s control mechanisms is provided in Table 1.

\begin{figure} [h]
  \centering
  \includegraphics[width=\linewidth]{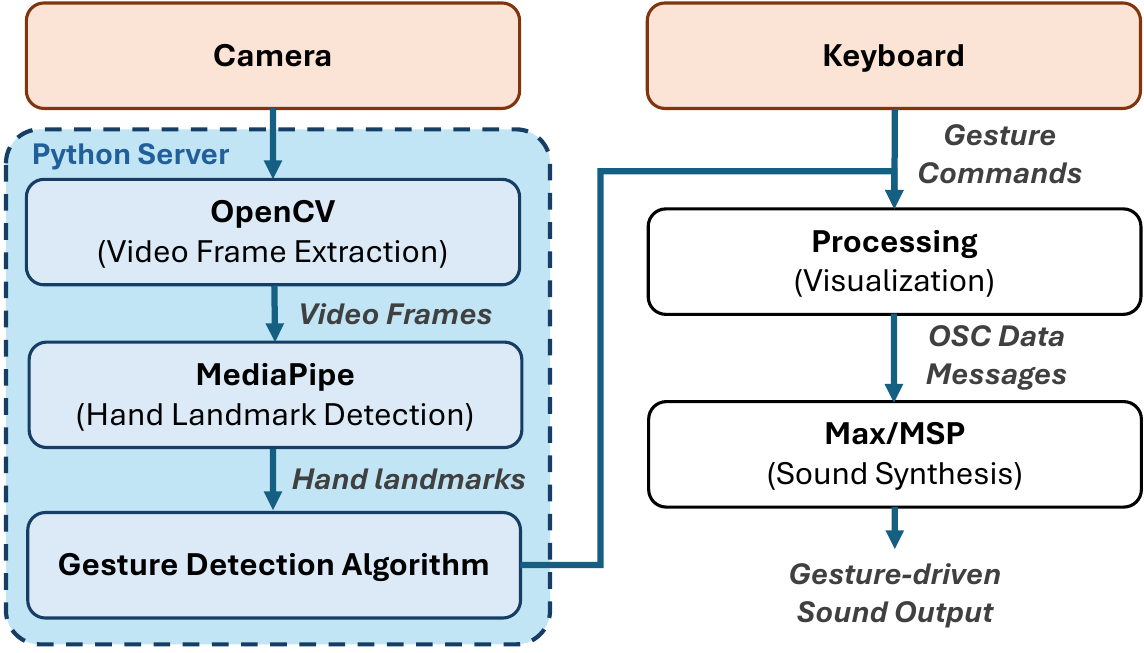}
  \caption{System architecture of \textit{NeoLightning}}
\end{figure}

The output is further shaped using \texttt{svf$\sim$} objects, which dynamically adjust the cutoff frequency in response to real-time changes in the sphere's X-axis position. The X-axis controls the cutoff frequency of a low-pass filter: as the sphere moves left, the cutoff frequency decreases, resulting in a progressively darker timbre. The user can amplify this effect in real time by increasing the size of the spheres, which in turn raises the system’s volume.

Each oscillator chain incorporates \texttt{live.gain$\sim$} objects for precise volume control, allowing nuanced amplitude adjustments. To enhance spatial perception, the patch integrates reverb effects that dynamically respond to the distance of the spheres. As the spheres move farther away, the reverb decay time increases, simulating a larger room. Combined with the visuals provided by Processing, this system allows users to intuitively grasp these relationships rather than relying solely on their imagination.

As the spheres move closer and begin to overlap, their oscillators interact, producing unexpected, evolving sounds. This effect simulates controlled feedback but in a more musical and intentional manner. To further explore these emergent sonic textures, we integrated the \texttt{P\_4L}\footnote{\url{https://ricola.gumroad.com/l/P_4L}, last access date: May 15, 2025} synthesizer into our patch, enabling dynamic sound generation.

By operating within a 3D control environment, the system enables more natural and multidimensional gestures, expanding the possibilities for sound design and performance. This enhanced depth of interaction demonstrates the system’s ability to push the boundaries of gesture-based musical interfaces, achieving a level of expressivity and immersion beyond that of traditional systems.


\section{Conclusion} 
This work presents a modern reinterpretation of the Buchla Lightning, addressing its reliance on proprietary hardware by integrating technologies such as MediaPipe for gesture recognition and Max/MSP with Processing for audiovisual processing. The redesigned system enhances accessibility and usability while preserving the expressive potential of the original instrument. A complementary keyboard input mode further ensures flexibility for users without access to a camera. Future developments could focus on refining gesture recognition algorithms to capture more nuanced movements and enable multi-user interactions for collaborative performances. Additionally, integrating spatial audio and adapting the system for mobile and augmented reality platforms could broaden its applicability. By combining historical ingenuity with modern advancements, this work demonstrates the potential of reimagined gesture-based music systems to inspire innovation in interactive sound design and performance.


\bibliography{icmc2025_paper_template}

\end{document}